\documentclass[aps,pra,graphics,twocolumn,floatfix,notitlepage,superscriptaddress]{revtex4-2}
\usepackage[english]{babel}
\usepackage{textcomp,xcolor,natbib}
\usepackage{dcolumn}    
\usepackage{graphicx}
\usepackage{amsmath, bbm}
\usepackage{latexsym}
\usepackage{amsfonts}   
\usepackage{amssymb}
\usepackage{array}      
\usepackage{epsfig}
\usepackage{txfonts}
\usepackage{xcolor,braket}
\usepackage[colorlinks=true,linkcolor=blue,urlcolor=blue,citecolor=blue]{hyperref}
\usepackage[normalem]{ulem}
\usepackage{soul}
\usepackage{dsfont}

\usepackage{xfrac}  
\usepackage{relsize}

\usepackage{mathrsfs}

\usepackage{cancel}
\usepackage{enumitem}

\usepackage{xspace}

\catcode`\>=\active \def>{
\fontencoding{T1}\selectfont\symbol{62}\fontencoding{\encodingdefault}}

\newcommand{\tr}{\ensuremath{\operatorname{tr}}}
\newcommand{\sqjsd}{\ensuremath{\sqrt{\text{JSD}}}\xspace}
\renewcommand{\vec}[1]{\ensuremath{\boldsymbol{#1}}}
\newcommand{\1}{\mathbbm{1}}
\newcommand{\nm}{non-Markovianity\xspace}

\begin{document}

\title{Entropic and trace distance based measures of non-Markovianity}

\author{Federico Settimo}
\affiliation{Dipartimento di Fisica “Aldo Pontremoli”, Università degli Studi di Milano, via Celoria 16, 20133 Milan, Italy}

\author{Heinz-Peter Breuer}
\email{breuer@physik.uni-freiburg.de}
\affiliation{Physikalisches Institut, Universität Freiburg, Hermann-Herder-Straße 3, D-79104 Freiburg, Germany}
\affiliation{EUCOR Centre for Quantum Science and Quantum Computing,
Universität Freiburg, Hermann-Herder-Straße 3, D-79104 Freiburg, Germany}

\author{Bassano Vacchini}
\email{bassano.vacchini@mi.infn.it}
\affiliation{Dipartimento di Fisica “Aldo Pontremoli”, Università degli Studi di Milano, via Celoria 16, 20133 Milan, Italy}
\affiliation{Istituto Nazionale di Fisica Nucleare, Sezione di Milano, via Celoria 16, 20133 Milan, Italy}

\begin{abstract}
  We analyze and compare different measures for the degree of non-Markovianity in the dynamics of open quantum systems. These measures are based on the distinguishability of quantum states which is quantified, on the one hand, by the trace distance or, more generally, by the trace norm of the Helstrom matrix, and, on the other hand, {by entropic quantifiers:} the Jensen-Shannon divergence, the Holevo or the quantum skew divergence. We explicitly construct a qubit dynamics for which the trace norm based non-Markovianity measure is nonzero, while all the entropic measures turn out to be zero. This leads to the surprising conclusion that the non-Markovianity measure which employs the trace norm of the Helstrom matrix is strictly stronger than all entropic non-Markovianity measures.
\end{abstract}

\maketitle

\section{Introduction}
\label{sec:introduction}
The study of quantum non-Markovian dynamics involves the investigation of
the very notion of stochastic process in the quantum realm, as well as
the characterization of memory effects in open quantum system
dynamics \cite{Breuer2016a, Rivas-quantum-nm,Li2018a,Milz2021a}. Memory effects in the dynamics of a quantum system
interacting with an external environment can be uniquely traced back
to local retrieval of exchanged information in the approach
to non-Markovianity based on the non monotonic behavior in time of
distinguishability of quantum states. This strategy was
introduced in \cite{Breuer2009b} and validated for different distinguishability
quantifiers of quantum states. In particular, while the original
approach was focused on the trace distance, it was later put into evidence
that invariance under translations of this quantifier led to failure
in assessing memory features in certain dynamics \cite{Liu-nonunital}. To avoid
this difficulty, the trace norm of the Helstrom matrix was used as a
generalized trace distance also sensitive to translations \cite{Chruscinski2011measures,Wissmann2015a}. A crucial
feature associated to the trace norm of the Helstrom matrix is the
fact that its non-monotonicity in time is equivalent to lack of P divisibility
of the considered dynamics, provided the evolution is invertible as a
linear transformation. In such a way a direct relation could be
established between a divisibility and a distinguishability criterion.

More recently, entropic distinguishability quantifiers have also been
introduced and directly connected to the notion of non-Markovianity as
due to information backflow \cite{Megier2021a}. To this aim, suitable regularizations of
the quantum relative entropy have been considered, which, at variance
with the quantum relative entropy, remain finite for any pair of states, and allow to introduce triangle-like inequalities which connect
revivals of the quantifier to information backflow, even in the
absence of a true triangle inequality as for distances.  Furthermore, these
entropic quantifiers are also sensitive to translations.  In this
framework, a special role is played by the Jensen-Shannon divergence,
whose square root is a true distance \cite{Briet2009a,Sra,Virosztek}.

Given that entropic distinguishability quantifiers are
contractions under positive trace preserving maps which are not
necessarily completely positive, as it happens for the trace distance
and the trace norm of the Helstrom matrix, a natural question is the
role of P divisibility in this context. Importantly, we show by means
of an example that the Jensen-Shannon divergence, as well as the other
entropic quantifiers, might fail in detecting breaking of P
divisibility.

Our results imply that the non-Markovianity measure
employing the trace norm of the Helstrom matrix is strictly
stronger than all of the entropic non-Markovianity measures, leading to a
nonzero value even for dynamics for which the entropic measures are
zero, while the opposite cannot happen.

The paper is organized as follows. In
Sec.~\ref{sec:entr-trace-dist} we introduce and exemplify the general
framework for the treatment of non-Markovianity based on
distinguishability quantifiers, together with the associated measures.
In Sec.~\ref{sec:non-mark-divis} we outline the connection between
non-Markovianity and divisibility of the dynamics, and explore this
relationship in its dependence on the considered distinguishability
quantifier. In particular, we construct an example of non-P divisible
evolution whose non-Markovianity measure is zero according to entropic
quantifiers. We summarize and discuss the conclusions of our work in
Sec.~\ref{sec:conclusions-outlook}.

\section{Entropic and trace distance based distinguishability quantifiers}
\label{sec:entr-trace-dist}

Let us begin by introducing the general framework of non-Markovianity for the dynamics of open quantum systems.
The main aim is to compare the well-known measure of memory effects based on the trace distance with other measures using alternative different distinguishability quantifiers between quantum states, in particular those related to the quantum relative entropy.

\subsection{Trace distance and Helstrom matrix}
\label{sec:trace-dist-helstr}
In the framework of quantum information and statistics there are many different quantifiers of distinguishability between two quantum states $\rho$ and $\sigma$.
A very important one is given by the trace distance (TD) \cite{Heinosaari-Ziman}
\begin{equation}
  D(\rho,\sigma) = \frac 12 \lVert\rho-\sigma\rVert,
  \label{eq:td}
\end{equation}
where the trace norm of any trace-class operator $A$ is defined as $\lVert A\rVert = \tr\sqrt{A^\dagger A}$.
The TD is bounded, $0\le D(\rho,\sigma)\le 1$, with $D(\rho,\sigma)=0$ if and only if $\rho=\sigma$, and $D(\rho,\sigma)=1$ if and only if $\rho\perp\sigma$.
Additionally, the TD obeys the triangle inequality
\begin{equation}
  D(\rho,\sigma)\leqslant D(\rho,\tau) + D(\tau,\sigma),
  \label{eq:td_triang}
\end{equation}
is contractive under the action of any completely positive trace preserving (CPTP) map $\Lambda$, as well as of any positive trace preserving map  \cite{Ruskai-beyond}
\begin{equation}
  D(\Lambda\rho, \Lambda\sigma)\le D(\rho,\sigma),
  \label{eq:td-contractivity}
\end{equation}
and it is invariant under unitary and anti-unitary transformations \cite{Molnar-isometries}.
%The TD can be extended from the  set of quantum states to the whole set of trace class operators.
It is also invariant under translations, in the sense that
\begin{equation}
  D(\rho+A, \sigma+A) = D(\rho,\sigma)
  \label{eq:trasla}
\end{equation}
for any operator $A$.
This follows directly from the fact that the TD depends on the difference between its two arguments.

It is possible to give the TD an interpretation as the bias in favour of a correct identification between two quantum states, upon performing a single measurement.
Let us suppose that Alice prepares the state $\rho$ or $\sigma$, each with probability $\frac12$, and sends it to Bob; the TD is linked to Bob's maximal probability of correctly distinguishing between the two as \cite{Fuchs-cryptographic}
\begin{equation}
  P_{\text{dist}}(\rho,\sigma)=\frac12(1+D(\rho,\sigma)).
  \label{eq:td-discrimination}
\end{equation}
This feature, combined with the contractivity of the TD under CPTP maps \eqref{eq:td-contractivity}, tells us that CPTP maps cannot increase the probability of distinguishing between quantum states.

%\subsubsection{Helstrom matrix}
The idea of using the trace norm $\lVert\cdot\rVert$ to quantify the bias in favour of a correct identification can be generalised also to the case in which the two states $\rho$ and $\sigma$ are not prepared with the same a-priori probability.
In fact, if one supposes that Alice prepares $\rho$ with probability $p$ and $\sigma$ with probability $1-p$, then Bob's maximal probability of distinguishing between the two is given by \cite{Helstrom-quantum-detection}
\begin{equation}
  P_{\text{dist}}(\rho,\sigma)=\frac12(1+\lVert\Delta\rVert),
  \label{eq:helstrom-discrimination}
\end{equation}
where
\begin{equation}
  \label{eq:helstrom}
  \Delta = p\rho - (1-p)\sigma
\end{equation}
is known as the Helstrom matrix \cite{Helstrom-detection-theory}.
The trace norm of $\Delta$ represents the bias in favour of a correct identification and Eq.~\eqref{eq:helstrom-discrimination} reduces to \eqref{eq:td-discrimination} in the unbiased case $p=\tfrac12$.
The Helstrom matrix can be seen as a generalisation of the TD to generic ensembles $\set{(p,\rho), (1-p,\sigma)}$ and it inherits properties such as boundedness and contractivity from the TD.

\subsection{Jensen-Shannon and skew divergences}
\label{sec:jensen-shannon-skew}
The TD is not the only possible quantifier of distinguishability between quantum states.
A particularly interesting distinguishability quantifier is the relative entropy
\begin{equation}
  \label{eq:rel-ent}
  S(\rho,\sigma) =
  \begin{cases}
    \tr[\rho\log\rho-\rho\log\sigma] \quad &\text{if }\operatorname{supp}\rho\subseteq \operatorname{supp}\sigma\\
    \infty\quad &\text{otherwise}
  \end{cases},
\end{equation}
where we take the logarithm in base 2.
The relative entropy, just like the TD, is contractive under both CPTP maps and positive trace preserving maps \cite{Reeb2017a}.
However, as it is evident from the definition, it is not bounded.

The relative entropy can also be naturally associated to a distinguishability task.
In particular, let us suppose to be able to prepare and measure the states an arbitrarily large number $N$ of times. 
{The relative entropy $S(\rho,\sigma)$ represents the maximal asymptotic rate at which the probability of erroneously concluding that the state is $\rho$, when it is actually $\sigma$, decays with the size $N$ of the sample over which a measurement is performed, so that, for large enough $N$ the probability of correctly identifying the state is \cite{audenaert-qsd, Bengtsson-Zyczkowski, Hayashi2006}}
\begin{equation}
    P_{N,\text{dist}}(\rho,\sigma) = 1-e^{-NS(\rho,\sigma)}.
\end{equation}
Unboundedness here arises naturally: whenever $\operatorname{supp}\rho\not\subseteq \operatorname{supp}\sigma$, it is possible with certainty to distinguish $\rho$ from $\sigma$ with only a finite number of measurements and hence the rate is infinite.

\begin{figure}[t]
  \centering
  \includegraphics[width=0.9\linewidth]{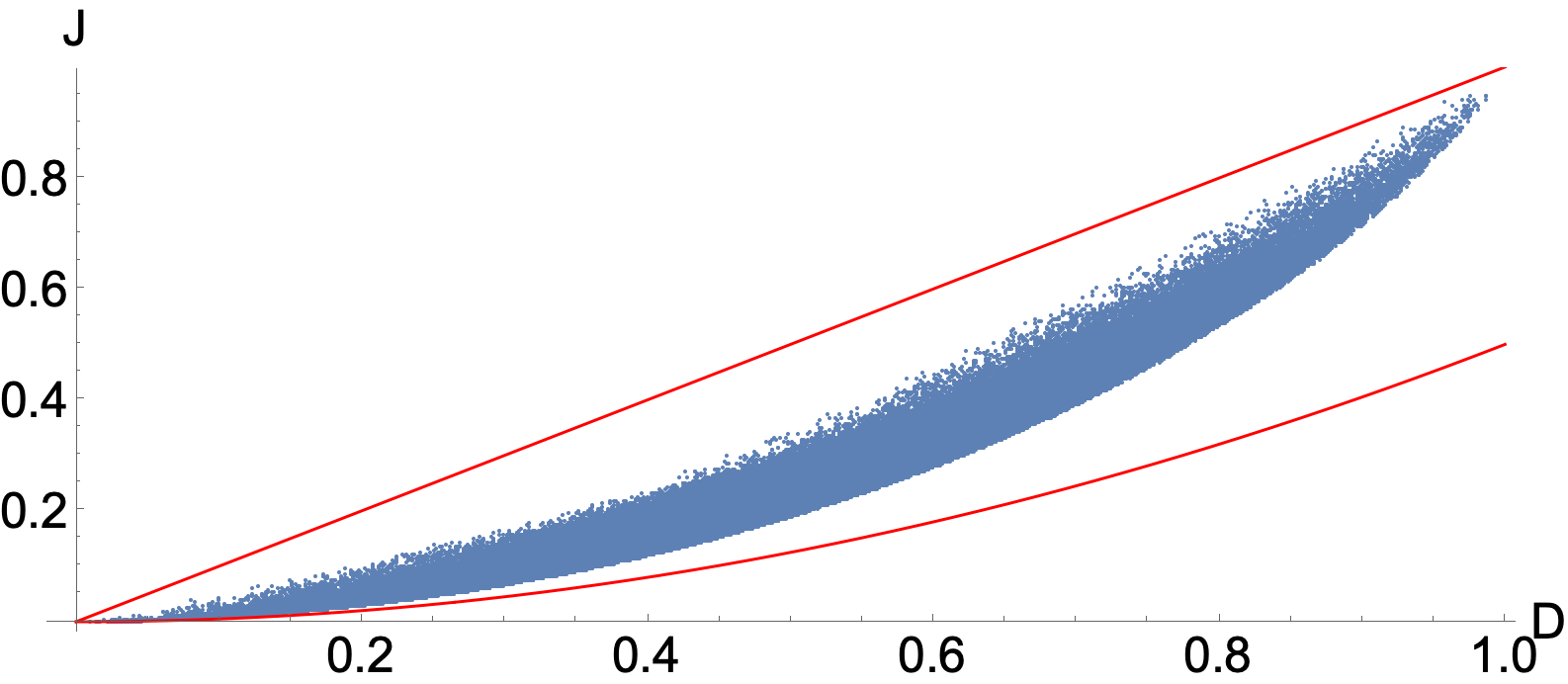}
  \caption{Plot of the JSD and the TD for $10^5$ randomly generated pairs of qubits. The red lines are the upper and lower bounds of Eq.~\eqref{eq:jsd-bounds}, which are valid also for arbitrarily dimensional Hilbert spaces.}
  \label{fig:bounds_J}
\end{figure}

\begin{figure}[t]
  \centering
  \includegraphics[width=\linewidth]{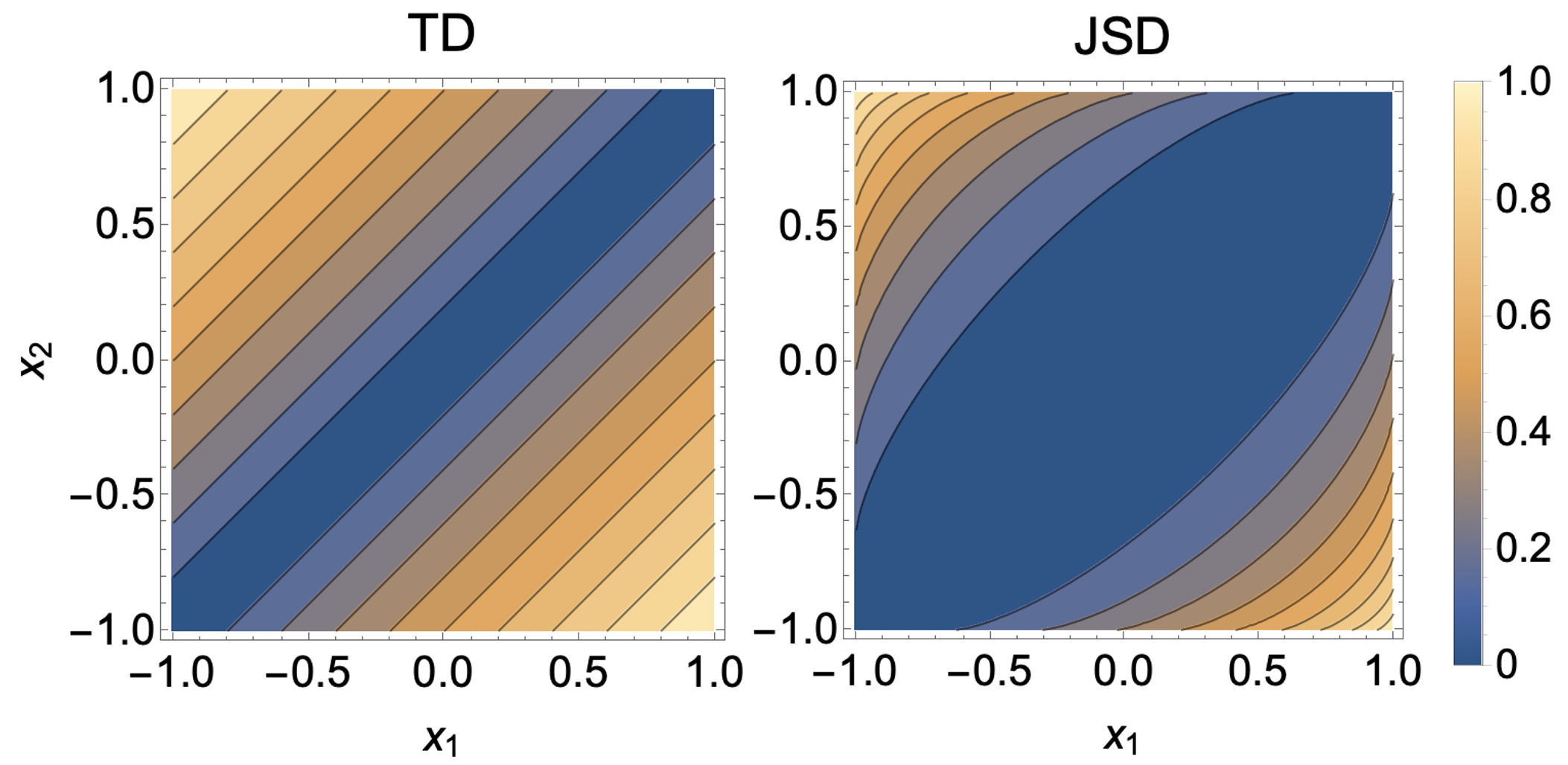}
  \caption{Plots of the TD (left) and of the JSD (right) for qubit states represented by Bloch vectors of the form $\vec r_\rho = (x_1,0,0)^\top$, $\vec r_\sigma = (x_2,0,0)^\top$. The translational invariance of the TD is reflected by the fact that the plot on the left only depends on the difference $x_1-x_2$. Note further the different sensitivity in the central and corner regions.}
  \label{fig:plots_J}
\end{figure}

It is possible to define a smoothed version of the relative entropy, namely the Jensen-Shannon divergence (JSD), according to \cite{Majtey-jsd}
\begin{equation}
  \begin{split}
    J(\rho,\sigma) =& \frac12 S\left(\rho,\frac{\rho+\sigma}2\right) + \frac12 S\left(\sigma,\frac{\rho+\sigma}2\right)\\
    =&H\left(\frac{\rho+\sigma}2\right) - \frac12 H(\rho) - \frac12 H(\sigma),
  \end{split}
  \label{eq:jsd}
\end{equation}
where $H$ denotes the von Neumann entropy $H(\rho)=-\tr \rho\log\rho$.
This definition ensures that the JSD inherits the contractivity under CPTP maps from the relative entropy and, additionally, it is bounded according to $0\le J(\rho,\sigma)\le 1$, with $J(\rho,\sigma)=0$ if and only if $\rho=\sigma$, while  $J(\rho,\sigma)=1$ if and only if $\rho\perp\sigma$.
In particular, it can be bounded by monotonic functions of the TD as \cite{Audenaert-tre-II, Pinsker}
\begin{equation}
  \label{eq:jsd-bounds}
  \frac12 D(\rho,\sigma)^2\le J(\rho,\sigma) \le D(\rho,\sigma).
\end{equation}  
The lower bound directly follows from the Pinsker inequality \cite{Bengtsson-Zyczkowski}.
Fig.~\ref{fig:bounds_J} shows these bounds 
together with the value of the TD and the JSD for randomly chosen pairs of states.
The JSD is not invariant under translations in the sense of \eqref{eq:trasla}, since, unlike the TD, it does not depend solely on the difference $\rho-\sigma$.
This fact is visualized in Fig.~\ref{fig:plots_J}.

The JSD, unlike the TD, is not a distance since it does not obey the triangle inequality.
However, it has been proven that its square root (\sqjsd) does obey this inequality and is indeed a distance \cite{Briet2009a,Sra, Virosztek}.
Even if it does not obeys the triangle inequality, the JSD obeys a {triangle-like} inequality
\begin{equation}
  \label{eq:jsd-triang}
  \begin{split}
  J(\rho,\sigma)-J(\rho,\tau)&\le \frac{1+D(\sigma,\tau)}2 - J(1,D(\sigma,\tau))\\
  &\leq 
  \sqrt[4]{2}\sqrt[4]{J(\sigma,\tau)},
  \end{split}
\end{equation}
which follows from the triangle-like inequalities presented in \cite{Audenaert-tre-II}.

It is possible to generalise the JSD to generic ensembles $\set{(\mu,\rho), (1-\mu,\sigma)}$, however, unlike for the TD, such generalisation is not unique.
As suggested in \cite{Smirne2022b} we point to two distinct generalizations based on a skewed version of the relative entropy, also called telescopic relative entropy in the quantum setting \cite{Audenaert-tre, Lee}. We therefore introduce the 
Holevo skew divergence 
\begin{equation}
  \label{eq:chi}
  % K_\mu(\rho,\sigma) = \frac1{h(\mu)}[H(\mu\rho+(1-\mu)\sigma)-\mu H(\rho) - (1-\mu)H(\sigma)],
  K_\mu(\rho,\sigma) = \frac{\chi_\mu(\rho,\sigma)}{h(\mu)},
\end{equation}
where
\begin{equation}
  \label{eq:binary-entropy}
  h(p) = -p \log p - (1-p)\log(1-p)
\end{equation}
is the binary entropy for the distribution $\set{p,1-p}$ and
\begin{equation}
  \label{eq:chi-Holevo}
  \chi_\mu(\rho,\sigma) = H(\mu\rho+(1-\mu)\sigma)-\mu H(\rho) - (1-\mu)H(\sigma)
\end{equation}
is the Holevo $\chi$ quantity \cite{Holevo-chi} for the considered ensemble, as well as the quantum skew divergence
\begin{equation}
\begin{split}
    S_{\mu} (\rho, \sigma) = & \frac{\mu}{\log (1 / \mu)} S (\rho, \mu \rho +
  (1 - \mu) \sigma) \\
  & + \frac{1 - \mu}{\log (1 / (1 - \mu))} S (\sigma, (1 - \mu) \sigma +
  \mu \rho).
\end{split}
    \label{eq:tre}
\end{equation}
Both quantities are bounded and they reduce to the JSD in the unbiased case $\mu=\frac12$.
Furthermore, they both obey triangle-like inequalities similar to the ones that hold for the unbiased case \eqref{eq:jsd-triang}, namely \cite{Megier2021a, Smirne2022b}
\begin{gather}
  \label{eq:skew-triang}
   S_\mu(\rho,\sigma)-S_\mu(\rho,\tau)\le \eta_\mu^S \sqrt[4]{S_\mu(\sigma,\tau)}
  \\
  \label{eq:holevo-triang}
  K_\mu(\rho,\sigma)-K_\mu(\rho,\tau)\le \eta_\mu^K\sqrt[4]{K_\mu(\sigma,\tau)},
\end{gather}
with
\begin{gather}
  \notag \eta_\mu^S = \log \left( \frac{1}{\mu (1 - \mu)} \right)\sqrt[4]{\frac{\mu (1 - \mu)}{2\, h (\mu) \log^3  (\mu) \log^3  (1 - \mu)}}, \\
  \eta_\mu^K = \sqrt[4]{\frac{8\mu(1-\mu)}{h(\mu)^3}}.
\end{gather}

\subsection{Non-Markovianity measures from distinguishability quantifiers}
\label{sec:non-mark-meas}
The unavoidable interaction between a quantum system and its surroundings leads to system-environment correlations and non-unitary time evolution of the state.
Assuming that at the initial time $t=0$ the global system-environment state is factorised, the dynamics is  described by a one-parameter family of CPTP dynamical maps $\Phi=\set{\Phi_t | 0\le t\le T, \Phi_0=\mathbbm1}$ such that $\rho(t) = \Phi_t\rho(0)$.
{Assuming that $\Phi_t^{-1}$ exists at all times $t\ge 0$, it is possible to define a two-parameter family of maps as
\begin{equation}
    \Phi_{t,s} = \Phi_t\Phi_s^{-1},\quad t\ge s\ge 0,
\end{equation}
such that $\Phi_{t,0}=\Phi_t$, describing the evolution of the state from time $s$ to time $t$.
The dynamics is said to be (C)P divisible if $\Phi_{t,s}$ is (completely) positive for all times $ t\ge s\ge 0$.}

The interaction between the system and the environment can lead to memory effects during the dynamics of the state.
If this happens, the dynamics is said to be non-Markovian.
Following \cite{Breuer2009b,Breuer2016a} we define a family of non-Markovianity measures, based on some distinguishability quantifier $d$, as
\begin{equation}
  \label{eq:nm-generic}
  \mathscr N^d(\Phi) = \sup\int\limits_{\sigma_d(t)>0}dt \, \sigma_d(t),
\end{equation}
where
\begin{equation}
  \sigma_d(t) = \frac d{dt}\, d(\rho_1(t),\rho_2(t)),
\end{equation}
and the maximisation is performed over all possible pairs of initial states $\rho_{1,2}(0)$ and any eventual parameter defining the distinguishability quantifier $d$, such as the skewing parameter $\mu$ defining $S_\mu$ or $K_\mu$.
A certain pair of initial states $\rho_{1,2}(0)$ is said to be optimal, if the maximum of equation \eqref{eq:nm-generic} is attained on this pair.
{Thus, a dynamical map $\Phi$ is Markovian according to the quantifier $d$ if and only if $\mathscr N^d(\Phi)=0$ or, equivalently, if $d(\rho_1(t), \rho_2(t))$ is a monotonic function of time for any initial pair of states $\rho_{1,2}(0)$.
Alternative approaches are indeed possible, such as violations of divisibility of the dynamical map \cite{Breuer2016a, Rivas-quantum-nm,Li2018a,Pollock2018a,Budini2018b,Budini2022a}.
}

Following \cite{Smirne2022b}, in order to have a well-defined measure of non-Markovianity we ask the quantifier $d$ to obey three properties:
\begin{enumerate}
\item Boundedness and indistinguishability of identical states:
  \begin{equation}
    \label{eq:boundedness}
    0\le d(\rho,\sigma)\le 1,
  \end{equation}
  with $d(\rho,\sigma) = 1$ if and only if $\rho\perp\sigma$, and $d(\rho,\sigma) = 0$ if and only if $\rho=\sigma$.
  Considering bounded distinguishability quantifiers allows to perform the maximation in Eq.~\eqref{eq:nm-generic} thus warranting that the measure of non-Markovianity is well-defined.
\item Contractivity under CPTP maps:
  \begin{equation}
    \label{eq:contractivity}
    d(\Lambda\rho, \Lambda\sigma)\le d(\rho,\sigma)
  \end{equation}
  for any CPTP map $\Lambda$.
  This property is crucial so that any revival in $d$ must necessarily correspond to violations of divisibility of the dynamical map.
  In fact, if $\Phi$ is CP divisible, $d$ must be monotonically decreasing, since the map $\Phi_{t,s}$ describing the evolution from $s$ to $t>s$ is always CPTP.
  Therefore, a revival in $d$ is possible only if $\Phi$ violates the divisibility.
  
\item Triangle-like inequalities:
  \begin{gather}
    \label{eq:triangle-like}
    d(\rho,\sigma)-d(\rho,\tau)\le\phi(d(\sigma,\tau)),\\
    d(\rho,\sigma)-d(\tau,\sigma)\le\phi(d(\rho,\tau)),
  \end{gather}
  where $\phi(x)$ is a strictly positive concave function for $x>0$, and with $\phi(0)=0$.
  This property allows for a microscopic interpretation of the revivals of $d$ as a twofold exchange of information, which is at first stored in external degrees of freedom and later retrieved in the open system.
\end{enumerate}
TD, JSD, \sqjsd, and their generalisations all obey properties 1-3, and hence lead to a well-defined measure of non-Markovianity. For the TD and the \sqjsd, which are actually distances, the function $\phi$ is given by the identity, while, for the JSD and the other entropic quantities, the function $\phi$ is proportional to the fourth root as follows from Eqs.~\eqref{eq:jsd-triang}, as well as \eqref{eq:skew-triang} and \eqref{eq:holevo-triang}.

{Given two distinguishability quantifiers $d_1$ and $d_2$ satisfying 1-3, we say that $\mathscr N^{d_1}$ is stronger than $\mathscr N^{d_2}$ if, for any dynamical map $\Phi$ such that $\mathscr N^{d_2}(\Phi)>0$, then $\mathscr N^{d_1}(\Phi)>0$.
Furthermore, $\mathscr N^{d_1}$ is strictly stronger than $\mathscr N^{d_2}$ if it is stronger and there exists $\Phi$ such that $\mathscr N^{d_2}(\Phi)=0$ and $\mathscr N^{d_1}(\Phi)>0$.
Viceversa, $\mathscr N^{d_1}$ is (strictly) weaker than $\mathscr N^{d_2}$ if $\mathscr N^{d_2}$ is (strictly) stronger than $\mathscr N^{d_1}$.
Two measures are said to be equivalent if $\mathscr N^{d_1}$ is both stronger and weaker than $\mathscr N^{d_2}$.}

An important distinguishability quantifier obeying the abovementioned three properties is the TD.
Optimal pairs for this measure must always be orthogonal and therefore on the border of the set of states \cite{Wissmann-optimal}.
Additionally, the {triangle inequality \eqref{eq:td_triang} allows to upper bound the revival of the TD from $s$ to a later time $t>s>0$ as \cite{Laine-witness,Amato2018a,Campbell2019b}}
\begin{equation}
  \label{eq:td-upperbound}
  \begin{split}
    \Delta D(t,s) =&D(\rho_{S}^1(t), \rho^2_{S}(t))-D(\rho_{S}^1(s), \rho^2_{S}(s))\\
    \le& D(\rho^1_{SE}(s), \rho^1_S(s)\otimes\rho^1_E(s))\\
    &+D(\rho^2_{SE}(s), \rho^2_S(s)\otimes\rho^2_E(s))\\
    &+D(\rho^1_E(s), \rho^2_E(s)),
  \end{split}
\end{equation}
where $\rho^i_{SE}(s)$, for $i=1,2$, is the global system-environment state, and $\rho^i_S (s)=\tr_E\rho^i_{SE}(s)$ and $\rho^i_E (s)=\tr_S\rho^i_{SE}(s)$ are, respectively, the reduced system and environmental states at time $s$.
This allows for a microscopic interpretation of the measure of non-Markovianity \cite{Laine-witness,Amato2018a}: a revival in the TD is possible only if at time $s$ the two environments are different or if correlations have built up during the dynamics.
Therefore, information is stored as correlations or as difference between the environmental states and can later flow back into the open system.
A similar interpretation also holds for the measure of non-Markovianity arising from the Helstrom matrix \cite{Wissmann2015a}, as well as for entropic distinguishability quantifiers \cite{Megier2021a,Smirne2022b}.
In particular, the non-Markovianity measure obtained according to Eq.~\eqref{eq:nm-generic} when the quantifier $d$ is the trace norm of the Helstrom matrix Eq.~\eqref{eq:helstrom}, which we denote as $\mathscr N^\Delta(\Phi)$, is positive if and only if $\Phi$ is not P divisible as has been shown in \cite{Chruscinski2011measures,Wissmann2015a} building on results in \cite{Kossakowski-necessary,Kossakowski-on-quant-stat-mech}.
The measure based on the TD, instead, is strictly weaker than $\mathscr N^\Delta$, since it can equal zero even for non-P divisible dynamics, due to its translational invariance \cite{Liu-nonunital}.
Given that both the TD and the quantum relative entropy are contractive under positive trace preserving maps, and the equivalence between a \nm measure and a divisibility property was obtained considering positivity, from now on we will concentrate our attention simply on positivity.

Let us now focus our attention to the entropic distinguishability quantifiers of Sec.~\ref{sec:trace-dist-helstr}.
Except for the relative entropy, which is unbounded, all the other quantifiers obey properties 1-3 and hence can be used to define a measure of non-Markovianity.
{In particular, we want to investigate whether these measures of non-Markovianity are equivalent to $\mathscr N^\Delta$.}
Namely, we want to know whether the equivalence between positivity of $\mathscr N^d(\Phi)$ and lack of P divisibility of $\Phi$ also holds when choosing $d$ as one of the previously introduced entropic quantifiers.
{We will show in Sec.~\ref{sec:counterexample} that this is not the case: we will use a counterexample to point out that $\mathscr N^\Delta$ is strictly stronger.
Let us focus in particular on the JSD, since the \sqjsd is just a monotonic function of it and hence $\mathscr N^ J$ and $\mathscr N^{\sqrt J}$ are equivalent.}

\subsection{Behavior on unital models}
\label{sec:behav-unit-models}
{Let us now focus our discussion on qubits, since it suffices considering the simplest non-trivial case to prove that $\mathscr N^\Delta$ is strictly stronger than $\mathscr N^J$.
For qubits, a generic state $\rho$ can be represented by means of a real three-dimensional Bloch vector $\vec r_\rho$ with $|\vec r_\rho|\le 1$ in the form
\begin{equation}
    \rho = \frac12(\1+\vec r_\rho\cdot \vec \sigma),
\end{equation}
where $\vec\sigma=(\sigma_x,\sigma_y,\sigma_z)^\top$ is the vector of the Pauli matrices.}
Under the action of a generic trace and Hermiticity preserving linear map, the Bloch vector associated to the state transforms according to
\begin{equation}
  \label{eq:representation-map-qubit}
  \vec r \mapsto \vec r_t = D(t)\vec r + \vec\kappa(t),
\end{equation}
where $D(t) = \operatorname{diag}\{\lambda_1(t), \lambda_2(t), \lambda_3(t)\}$ is a real diagonal $3\times3$ matrix and $\vec\kappa(t)$ is a real three-dimensional vector \cite{King-minimal}. The representation Eq.~\eqref{eq:representation-map-qubit} is valid up to orthogonal transformations which in the present context plays no role, given that both TD and JSD are invariant under unitary transformations, and the measure of non-Markovianity in Eq.~\eqref{eq:nm-generic} is obtained by maximizing over the possible initial states.
Let us focus in particular on unital maps, which are the maps that preserve the maximally mixed state at any time $t\ge0$: $\Phi_t\left[\frac\12\right]=\frac\12$.
Alternatively, employing the representation \eqref{eq:representation-map-qubit}, they are the maps such that $\vec\kappa(t)=0$ at all times $t\ge0$.
A dynamics of this kind is not P divisible if and only if at least one of the functions $\lambda_i(t)$ does not decrease monotonically.

An important feature of unital dynamics is that $\mathscr N^D(\Phi)>0$ if and only if $\Phi$ violates P divisibility: any backflow of information, corresponding to a violation of P divisibility, is witnessed by the TD, without the need to generalise it to the Helstrom matrix.
Interestingly, this feature also holds for the JSD.
{Let $\rho_{1,2}(0)$ be the optimal pair for the TD.  Since
  they must be pure and orthogonal, we have $\rho_1(0)+\rho_2(0) = \1$
  and the Bloch vectors representing the states obey
  $\vec r_1(0)=-\vec r_2(0)$.  Thanks to unitality the transformed average state
  $(\rho_1(t)+\rho_2(t))/2$ remains the maximally mixed
  state, so that $\vec r_1(t)=-\vec r_2(t)$ holds at all times. Thus the TD
  between the two states reads $D(\rho_1(t),\rho_2(t))=r(t)$ and
  both evolved states have the same von Neumann entropy
\begin{equation}
    H(\rho_1(t)) = H(\rho_2(t)) = h\left(\frac{1-r(t)}2\right),
\end{equation}
where $h$ is the binary entropy introduced in Eq.~\eqref{eq:binary-entropy}.
It is therefore possible to rewrite the JSD using Eq.~\eqref{eq:jsd} as
\begin{equation}
  \label{eq:J-D-opt-pairs}
      J(\rho_1(t), \rho_2(t)) = 1-h\left(\frac{1-D(\rho_1(t), \rho_2(t))}2\right).
\end{equation}
This expression is a monotonic function of the TD and thus a revival in the JSD is witnessed if and only if it is witnessed by the TD.
Therefore, as it happens for the TD, $\mathscr N^J(\Phi)>0$ if and only if $\Phi$ violates P divisibility.}
Unlike the TD, the characterisation of optimal pairs for the JSD is still an open problem.
For unital maps acting on qubits, numerical evidence suggests that they must be pure and orthogonal, just like for the TD.
This feature allows for an interesting interpretation of $\mathscr N^J$ in terms of the von Neumann entropy.
By employing the first line of Eq.~\eqref{eq:J-D-opt-pairs}, holding for any pair of pure and orthogonal initial states, it is therefore possible to rewrite the measure of \nm as
\begin{equation}
  \label{eq:N-J-unital}
  \mathscr N^J(\Phi) = \max_{\rho(0)}\int_{\Gamma_{\rho(0)}}dt \frac d{dt}[-H(\rho(t))],
\end{equation}
where $\Gamma_{\rho(0)} = \set{t\in\mathbb R|\frac d{dt}H(\rho(t))<0}$.
The measure of \nm for the JSD in the case of unital dynamics is given by the total decrease of entropy for a single state, maximised over all possible initial states.
This is clearly linked to violations of P divisibility of $\Phi$, since any unital positive map acting on qubits increases the entropy of the state \cite{Bengtsson-Zyczkowski}, so that any revival in the entropy must necessarily correspond to a violation of P divisibility.
Unfortunately, this feature is only true for qubits, since in higher dimensions orthogonal states do not need to have the same eigenvalues.
Additionally, no similar interpretation holds for \sqjsd or for the two generalisations to ensembles $K_\mu$ and $S_\mu$.

\subsection{Robustness of optimal pairs}
\label{sec:robustn-optim-pairs}
Let us now study the robustness of optimal pairs, i.e. how the measure of \nm changes when moving away from the optimal pair, for the different distinguishability quantifiers.
We will illustrate this by considering a simple but yet paradigmatic model: the dephasing model.
This model consists in a modification of the coherences without a corresponding change in the populations:
\begin{equation}
  \label{eq:dephasing}
  \rho(t) =
  \begin{pmatrix}
    \rho_{00}&\rho_{01}\gamma(t)e^{-i\omega_St}\\
    \rho_{10}\gamma^*(t) e^{i\omega_St}&\rho_{11}
  \end{pmatrix},
\end{equation}
where $\gamma$ is called the decoherence function.
For this model, \nm corresponds to a non-monotonic behaviour of $\lvert\gamma\rvert$.
It is worth stressing that considerations similar to the ones for this model, are also valid for other models such as, for example, the phase covariant model which will be introduced in Sec.~\ref{sec:phase-covar-dynam}.

\begin{figure}[t]
\centering
\includegraphics[width=\linewidth]{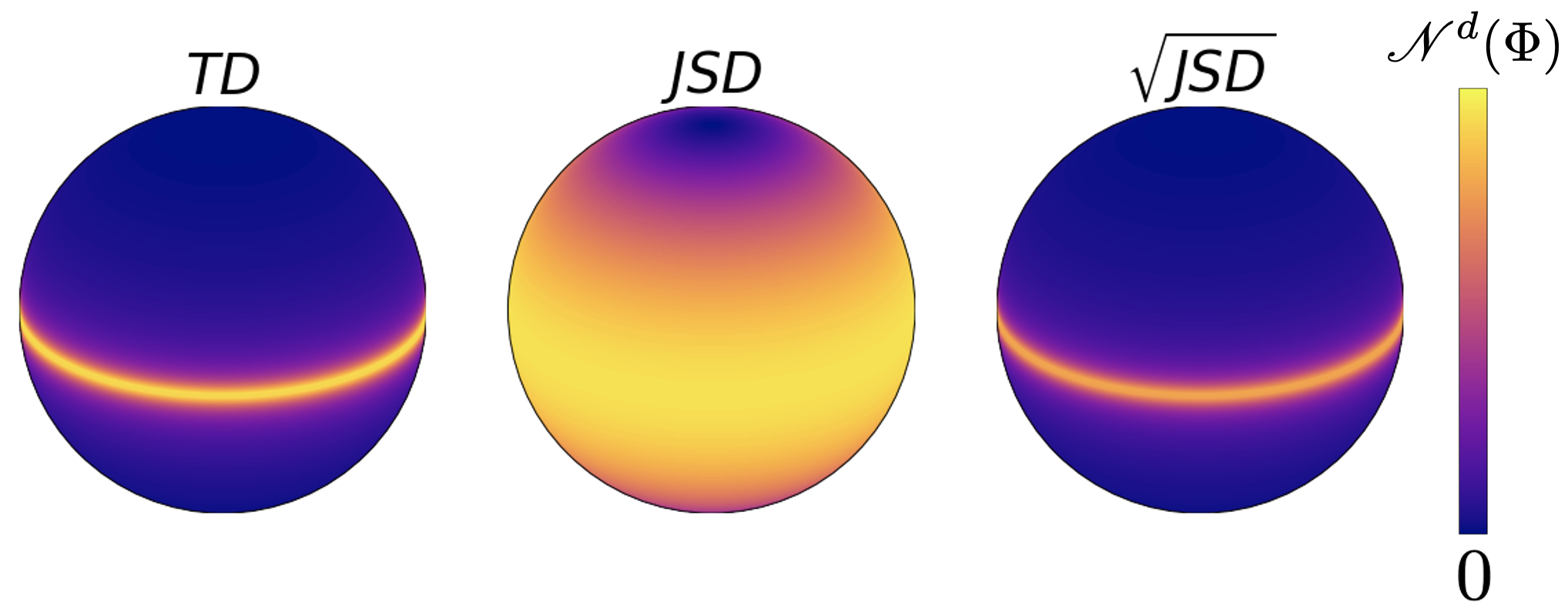}
\caption{{Measure of non-Markovianity for pure and orthogonal
states for TD (left), JSD (middle), and \sqjsd (right).
The color provide the value of the measure of non-Markovianity, rescaled according to the maximum value reached by the measure for the considered distinguishability quantifier $d$, obtained considering as initial states the corresponding point on the surface of the Bloch sphere and its antipodal point.
Brighter colors correspond to higher values of the revivals.
The reference values are taken to be $1\times 10^{-2}$ for $\mathscr N^D(\Phi)$ as well as $\mathscr N^{\sqrt J}(\Phi)$, and $1\times 10^{-3}$ for $\mathscr N^J(\Phi)$, reflecting the different scales of the revivals.}
The two distances (TD and \sqjsd) behave very similarly.
For the JSD, on the other hand, the value of the measure of \nm decreases more slowly when moving away from the optimal pair, i.e. all the pairs of pure and orthogonal vectors on the equator of the Bloch sphere. The dynamics $\Phi$ is given by the dephasing model of Eq.~\eqref{eq:dephasing}
with decoherence function $\gamma(t)$ corresponding to a bosonic bath described by a spectral density with an exponential cut-off of the form $J(\omega)=\lambda({\omega^s}/{\Omega^{s-1}})\exp({-{\omega}/{\Omega}})$ as considered in \cite{Addis2014a,Guarnieri-quantum}, with $s=3$, $\lambda=3$ and $\Omega=1$ in inverse units of time.
}
\label{fig:stability-N-deph}
\end{figure}

Optimal pairs are all the pairs of pure and orthogonal states corresponding to antipodal vectors on the equator of the Bloch sphere, since the $x$ and $y$ direction are the only ones in which the dynamics is not trivial.
In order to evaluate the robustness of the optimal pairs, Fig.~\ref{fig:stability-N-deph} shows the behavior of the measure of \nm when moving away from the equatorial plane, but still considering pure and orthogonal states.
It is possible to notice that the TD and the \sqjsd, being both distances, behave very similarly, with the maximum of the measure of \nm on the equatorial plane, and quickly decreasing when moving towards the poles of the Bloch sphere.
For the JSD, on the other hand, the situation is qualitatively different, with a broader region around the equator with a measured value of \nm similar to the maximal one, which is about one order of magnitude smaller than the value obtained for the other quantifiers.

\section{Non-Markovianity and divisibility}
\label{sec:non-mark-divis}
In the definition of a measure of \nm for a generic distinguishability quantifier $d$, the condition of contractivity \eqref{eq:contractivity} implies that every P divisible dynamics leads to a zero measure of \nm.
On the other hand, using the Helstrom matrix, as soon as the dynamics violates P divisibility one has a non-zero measure of \nm, $\mathscr N^\Delta(\Phi)>0$.
{In other words, $\mathscr N^\Delta$ is stronger than $\mathscr N^d$ for any other quantifier $d$.
We now want to study whether there exists other quantifiers $d$ leading to measures that are equivalent to the one arising from the Helstrom matrix, with particular focus on the entropic quantifiers.

We already know that the properties 1-3 are not sufficient in order to have a measure of \nm equivalent to $\mathscr N^\Delta$, since it is strictly stronger than $\mathscr N^D$.
In Sec.~\ref{sec:counterexample} we will show with a counterexample that the same also holds for the JSD and its generalisations.
}

\subsection{Positivity and non-contractivity domain}
\label{sec:posit-non-contr}
Let us first tackle the question of the behavior of the JSD under non positive maps.
We already know that the JSD is contractive under any positive map.
We now want to investigate if the reverse is also true, namely we want to clarify whether, for any non positive map $\Lambda$, there exists a pair of states for which the JSD is strictly non contractive.
Non-positivity of $\Lambda$ implies that there exists some state $\rho$ which is mapped to a non-positive operator $\Lambda\rho$.
However, the JSD, unlike the TD, cannot be extended to non-positive operators, since it involves the logarithm of the eigenvalues.
Therefore, the search for a non-contractive pair for $\Lambda$ must be restricted to the set of states that are mapped to states after the action of the map, i.e. to the positivity domain

\begin{equation}
  \label{eq:pos-dom}
  \mathcal{PD}_\Lambda = \set{\rho\in\mathcal S(\mathscr H)|\Lambda\rho\in\mathcal S(\mathscr H)},
\end{equation}
where $\mathcal S(\mathscr H)$ is the set of quantum states on a Hilbert space $\mathscr H$.
In the following, we will only consider qubits $\mathscr H = \mathbb C^2$, since this will turn out to be sufficient in order to show that $\mathscr N^J$ is strictly weaker than $\mathscr N^\Delta$.
We denote the set of all qubit states, i.e. the Bloch sphere, as $\mathcal S(\mathbb C^2) = \mathcal S$.

Considering unital non-positive maps $\Lambda$, it is easy to show that there always exists a non-contractive pair inside $\mathcal{PD}_\Lambda$.
Such maps act on Bloch vectors according to Eq.~\eqref{eq:representation-map-qubit} with $\vec \kappa =0$ and non-positivity implies that some $\lambda_i>1$, which we take to be $\lambda_1$, without loss of generality.
The non-contractive pair is the one represented by the Bloch vectors $\vec r_\rho = (\lambda_1^{-1},0,0)^\top = -\vec r_\sigma$.
In fact, by direct calculation it is easy to show that
$J(\rho,\sigma) < J(\Lambda\rho,\Lambda\sigma) = 1$.
In the general case, an analytic proof for the existence of a non-contractive pair is missing.
However, by parameterising the non positive map $\Lambda$ as in Eq.~\eqref{eq:representation-map-qubit} and performing a sample on all the parameters, we observed numerically that for any such map it is always possible to find a pair of states $\rho,\sigma\in\mathcal{PD}_\Lambda$ such that $J(\Lambda\rho,\Lambda\sigma)>J(\rho,\sigma)$.

\begin{figure}[t]
  \centering
  \includegraphics[width=0.7\linewidth]{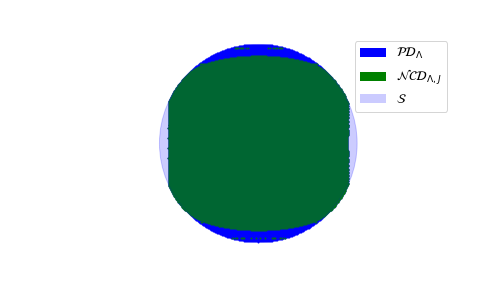}
  \caption{Section at $y=0$ of the Bloch sphere (light blue) for an example of a non-positive map $\Lambda$ for which the non-contractivity domain $\mathcal{NCD}_{\Lambda,J}$ (green) is strictly included in the positive domain $\mathcal{PD}_\Lambda$ (blue).
    This map acts on a Bloch vector $\vec r = (x,y,z)^\top$ as $\vec r \mapsto (\lambda_xx,\lambda_yy,\lambda_zz)^\top$, with $\lambda_x=\lambda_y=1.1$, and $\lambda_z=0.1$.}
  \label{fig:non-pos-dom}
\end{figure}

Turning back to the dynamical point of view, however, the search for the non-contractive pair might not be extended to all $\mathcal{PD}_\Lambda$.
In fact, not all the domain of positivity of $\Lambda=\Phi_{t,s}$ is available, but only the image at time $s$ of the Bloch sphere $\Phi_s(\mathcal S)$ is. We stress that $\Phi_s(\mathcal S)$ is in general only a subset of $\mathcal{PD}_\Lambda$.
Thus, in order to have $\mathscr N^J(\Phi)>0$ for all non-P divisible processes, we would need to be able to find a non-contractive pair for the JSD inside $\Phi_s(\mathcal S)$.
Let us now define the set of states in which it is possible to find a non-contractive pair as the non-contractivity domain
\begin{equation}
  \label{eq:non-cont-dom}
  \begin{split}
    \mathcal{NCD}_{\Lambda,J}=\{\rho\in&\mathcal{PD}_\Lambda\text{ }|\text{ }\exists\sigma\in\mathcal{PD}_\Lambda,\\
    &J(\Lambda\rho,\Lambda\sigma)>J(\rho,\sigma)\}.
  \end{split}
\end{equation}
Therefore, in order to have \nm for all non-P divisible dynamical maps, we would need to have $\mathcal{NCD}_{\Lambda,J}=\mathcal{PD}_\Lambda$: for any state in $\Phi_s(\mathcal S)\cap\mathcal{PD}_\Lambda$ it is always possible to find another state such that non-contractivity holds.
However, this is not the case, as it is clear from the example of Fig.~\ref{fig:non-pos-dom}.
There, in fact, {$\mathcal{NCD}_{\Lambda,J}$ is a proper subset of $\mathcal{PD}_\Lambda$}.
Therefore, if we were able to construct a dynamics $\Phi$ such that for times $t>s>0$ it acts as this non-positive map ($\Phi_{t,s}=\Lambda$), with a dynamics prior to time $s$ that is P divisible and with {the Bloch sphere that is mapped at time $s$ inside $\mathcal{PD}_\Lambda$ but outside $\mathcal{NCD}_{\Lambda,J}$, i.e. $\Phi_s(\mathcal S)\subset\mathcal{PD}_\Lambda \setminus \mathcal{NCD}_{\Lambda,J}$, we would construct a non-P divisible dynamics but with $\mathscr N^J(\Phi)=0$.}
This is indeed feasible as we will show in Sec.~\ref{sec:counterexample} providing explicitly  a model which is similar in spirit to the one just described.
%Let us first set the theoretical framework for the class of dynamics that will be used to construct this counterexample: the phase covariant dynamics.

\subsection{Phase covariant dynamics}
\label{sec:phase-covar-dynam}
In order to construct the counterexample of  Sec.~\ref{sec:counterexample}, let us first set the theoretical background of the considered dynamics, namely phase covariant ones.
They contain a  broad class of dynamics and they involve maps $\Phi$ that satisfy covariance with respect to phase transformations, namely \cite{Filippov2020b}
\begin{equation}
  \label{eq:phase-covariance}
  e^{-i\sigma_z\theta}\Phi_t[\rho]e^{i\sigma_z\theta} = \Phi_t\left[e^{-i\sigma_z\theta}\rho e^{i\sigma_z\theta}\right]
\end{equation}
for all real $\theta$ and for all states $\rho\in\mathcal S(\mathbb C^2)$.
Phase covariant dynamics are in the form \cite{Haase-fundamental, Smirne-ultimate}
\begin{equation}
  \label{eq:ph-cov-dyn}
  \Phi_t\rho = \frac12\big[\1 + \eta_\perp(t)(\nu_x\sigma_x+\nu_y\sigma_y) + \eta_\parallel(t)\nu_z\sigma_z + \kappa_z(t)\sigma_z\big],
\end{equation}
where $\nu_i = \tr[\rho\sigma_i]$, for $i=x,y,z$.
The complete positivity conditions reads
\begin{equation}
  \label{eq:ph-cov-CP}
  \eta_\parallel\pm\kappa_z\le1,\qquad
  1+\eta_\parallel\ge\sqrt{4\eta_\perp^2+\kappa_z^2}.
\end{equation}
The dynamics can be reformulated in terms of a master equation of the form
\begin{align}
  \label{eq:ph-cov-master-eq}
  \frac {d\rho}{dt} =& \gamma_+(t)\left(\sigma_+\rho\sigma_--\frac12\{\rho,\sigma_-\sigma_+\}\right)\\\notag
  &+\gamma_-(t)\left(\sigma_-\rho\sigma_+-\frac12\{\rho,\sigma_+\sigma_-\}\right)
  +\gamma_z(t)(\sigma_z\rho\sigma_z-\rho),
\end{align}
where
\begin{equation}
  \label{eq:ph-cov-rates}
  \gamma_\pm(t) = \frac{\eta_\parallel(t)}2\frac d{dt}\left(\frac{1\pm\kappa_z(t)}{\eta_\parallel(t)}\right),\quad
  \gamma_z(t) = \frac14\frac d{dt}\ln\frac{\eta_\parallel(t)}{\eta_\perp^2(t)}.
\end{equation}
The dynamics is CP divisible if and only if $\gamma_\pm(t)\ge0$ and $\gamma_z(t)\ge 0$.
P divisibility, instead, is satisfied whenever \cite{Filippov2020b}
\begin{equation}
  \label{eq:ph-cov-P-div}
  \gamma_\pm(t)\ge0
  \quad\text{and}\quad
  \sqrt{\gamma_+(t)\gamma_-(t)}+2\gamma_z(t)>0.
\end{equation}

The composition of two phase covariant dynamics is again phase covariant.
If we suppose that the system undergoes a first phase covariant dynamics $\Phi^1$ from $t=0$ to $t=t_1$, and later it evolves following $\Phi^2$, then the total dynamics $\Phi=\Phi^2\circ\Phi^1$, defined as
\begin{equation}
  \label{eq:ph-cov-composition}
  \Phi_t=
  \begin{cases}
    \Phi^1_t,\quad&\text{if }t\le t_1\\
    \Phi^2_{t-t_1}\Phi^1_{t_1},\quad&\text{if }t>t_1
  \end{cases},
\end{equation}
is again phase covariant, described by the functions
\begin{gather}
  \label{eq:ph-cov-comp-law-1}
  \eta_{\parallel,\perp}(t) =
  \begin{cases}
    \eta_{\parallel,\perp}^1(t)\quad&\text{if }t\le t_1\\
    \eta_{\parallel,\perp}^2(t-t_1) \eta_{\parallel,\perp}^1(t_1) \quad&\text{if }t>t_1
  \end{cases},\\\label{eq:ph-cov-comp-law-2}
  \kappa_z(t) =
   \begin{cases}
    \kappa^1_z(t)\quad&\text{if }t\le t_1\\
    \kappa^2_z(t-t_1)+\eta_\parallel^2(t-t_1)\kappa^1_z(t_1)\quad&\text{if }t>t_1
  \end{cases},
\end{gather}
where the superscripts 1 or 2 label the functions defining respectively $\Phi^1$ or $\Phi^2$. 
The composition of two phase covariant dynamics is not commutative, since in general $\Phi^1\circ\Phi^2\ne\Phi^2\circ\Phi^1$ as it is evident from \eqref{eq:ph-cov-comp-law-1} and \eqref{eq:ph-cov-comp-law-2}.
Furthermore, the family of all phase covariant dynamics do not form a group, since, in general, the inverse of a dynamics $\Phi^{-1}$ is not positive.

\subsection{Example showing that $\mathscr N^J$ is strictly weaker than $\mathscr N^\Delta$}
\label{sec:counterexample}
Let us now employ the previously introduced phase covariant model in order to build a counterexample of a dynamics which is not P divisible but yet leading to a zero measure of \nm for the JSD.
It follows from this counterexample that $\mathscr N^J$ is a strictly weaker measure of \nm than $\mathscr N^\Delta$, since $\mathscr N^\Delta>0$ for all non-P divisible dynamics.
We will actually consider a dynamics for which the memory effects are already detected by the TD, without the need to generalise to the Helstrom matrix.

\begin{figure}[t]
  \includegraphics[width=\linewidth]{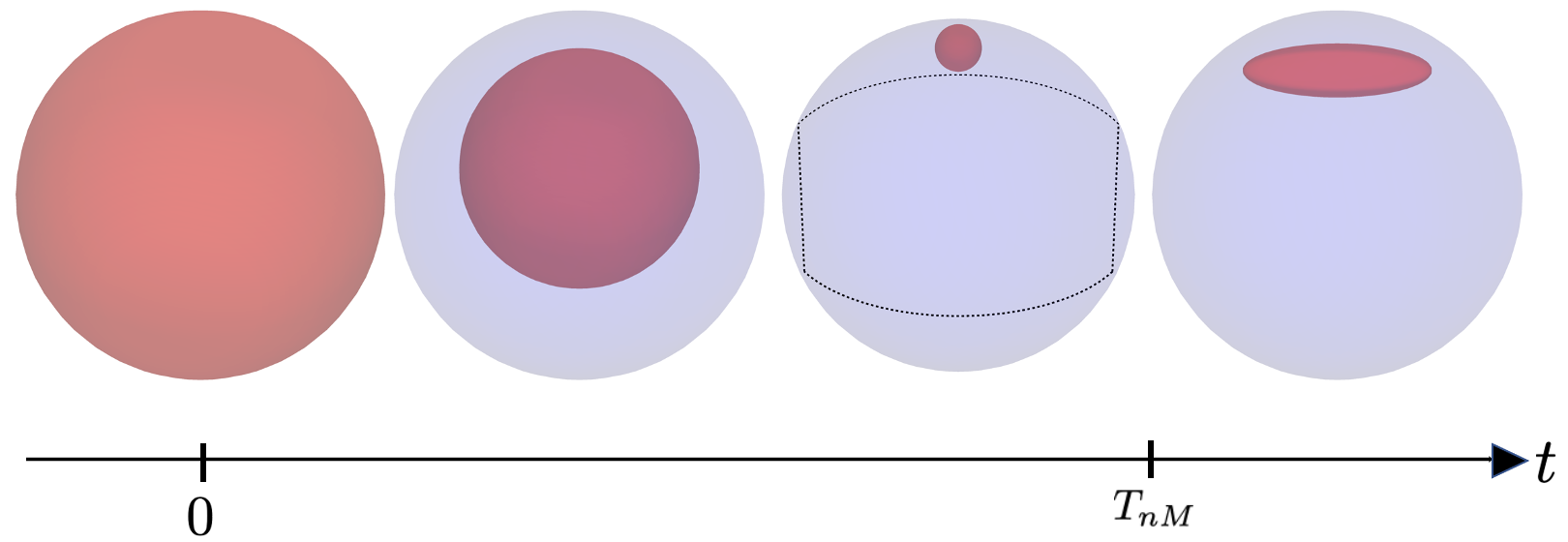}
  \caption{Schematic representation of the dynamics of the counterexample, obtained by visualizing the Bloch sphere (blue) and the evolved ellipsoid (red) at different time steps.
  {After time $t=0$ and till time $T_{nM}$ the image of the Bloch sphere is a uniformly contracted sphere translated in the $z$ direction, without violating P divisibility.
  After $T_{nM}$, the image of the sphere is further contracted and translated in the $z$ direction, but also expanded in the $x$ and $y$ directions, thus violating P divisibility.
  In the third panel, the dashed line represents the border of the non-contractivity domain for the dynamical map from $T_{nM}$ onwards, corresponding to the green area in Fig.~\ref{fig:non-pos-dom}.
  It thus clearly appears that the Bloch sphere at $T_{nM}$ is mapped outside the non-contractivity domain.}
  This way, even if the second part of the dynamics is not P divisible, there is no pair of states available for the JSD to witness a revival.}
  \label{fig:counter-representation}
\end{figure}

%However, this is not always the case: it is also possible to produce a counterexample of a non P divisible dynamics by considering a phase covariant dynamics with $\eta_{\parallel,\perp}$ monotonic and an oscillating translation $\kappa$ such that the dynamics violates P divisibility.
%This way, $\mathscr N^\Delta(\Phi)>0$, but $\mathscr N^D(\Phi)=0$ due to the translational invariance of the TD. Thus, $\mathscr N^D$ is strictly weaker than $\mathscr N^\Delta$.

%We show now with a counterexample that also $\mathscr N^J$ is strictly weaker than $\mathscr N^\Delta$.

\begin{figure*}[t]
  \includegraphics[width=0.45\linewidth]{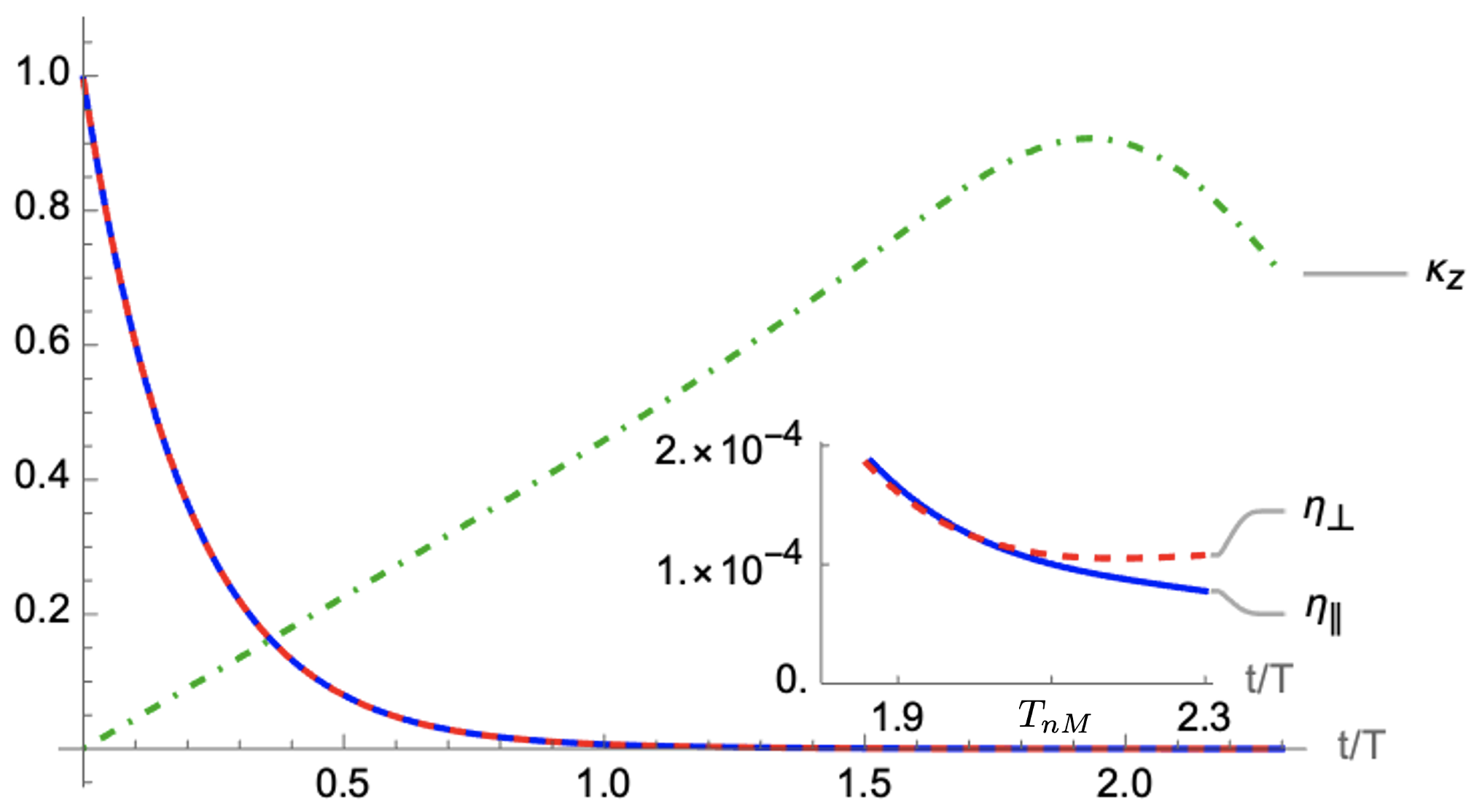}~~\includegraphics[width=0.45\linewidth]{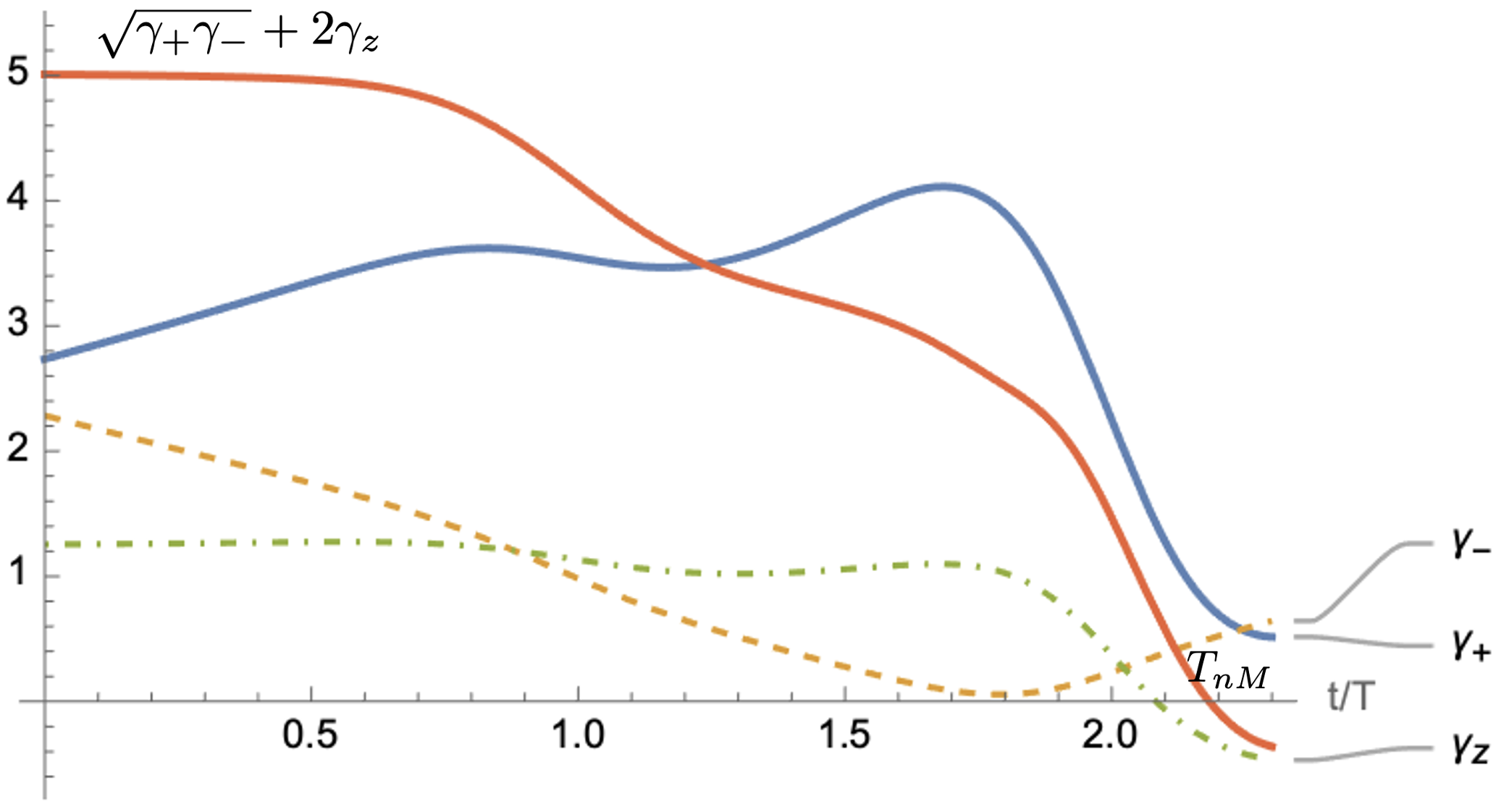}
  \caption{Left panel: The functions of Eqs.~\eqref{eq:couterexample-eta} and \eqref{eq:couterexample-kappa}.
  Non-Markovianity is present after $T_{nM}\approx 2.2 T$ and corresponds to a non-monotonic behavior of $\eta_\perp$.
  Right panel: Rates $\gamma_\pm$ and $\gamma_z$ for the corresponding dynamics.
  The red line corresponds to $\sqrt{\gamma_+\gamma_-}+2\gamma_z$, whose positivity is necessary for the dynamics to be P divisible, together with the conditions $\gamma_{\pm} > 0$.
    Non-Markovianity for $t>T_{nM}$ corresponds to a violation of divisibility. We considered  the choice of the parameters: $\mu_1 = 5$, $\mu_2=4$, $A_\parallel=0.01$, $A_\perp=1.01$, $A_\kappa = 0.45$ and $\alpha=5$.
  }
  \label{fig:counter-functions-rates}
\end{figure*}

The dynamical map $\Phi$ of such counterexample is described by the functions
\begin{gather}
  \label{eq:couterexample-eta}
  \begin{split}
    \eta_{\parallel,\perp}(\tau) =& e^{-\mu_1 \tau}\sigma(1-\tau) \\
    &+ e^{-\mu_1}e^{-\mu_2(\tau-1)}\sigma(\tau-1)\sigma(2-\tau) \\
    &+ e^{-\mu_1-\mu_2}\left[(3-\tau)+A_{\parallel,\perp}(\tau-2)\right]\sigma(\tau-2),
  \end{split}\\
  \label{eq:couterexample-kappa}
  \begin{split}
      \kappa_z(\tau) =& A_\kappa \tau \sigma(2-\tau) \\&+ 2 A_\kappa \left[(3-\tau)+A_{\parallel}(\tau-2)\right]\sigma(\tau-2),
  \end{split}
\end{gather}
where $\sigma$ is the sigmoid function
\[
  \sigma(\tau) = \frac1{1+e^{-\alpha \tau}},
\]
which is a smooth version of the Heaviside theta function and $\tau = t/T$ is a dimensionless time parameter, where $T$ is a reference time determining the duration of the different stages depicted in Fig.~\ref{fig:counter-representation}.
These functions together with the corresponding rates $\gamma_\pm$ and $\gamma_z$ obtained from Eq.~\eqref{eq:ph-cov-rates} are shown in Fig.~\ref{fig:counter-functions-rates}.

The idea of the counterexample follows from the considerations of Sec.~\ref{sec:posit-non-contr}: the dynamics in the time interval in which the memory effects arise will consist of a non-positive map $\Phi_{t,s}$ similar to the one described in Fig.~\ref{fig:non-pos-dom}, for which the non-contractivity domain \eqref{eq:non-cont-dom} is strictly smaller than the positivity domain \eqref{eq:pos-dom}.
Prior to this time interval, the dynamics is P divisible and such that it maps the whole Bloch sphere {inside  $\mathcal{PD}_{\Phi_{t,s}}$ but outside $\mathcal{NCD}_{\Phi_{t,s},J}$}, so that there is no pair of states available for the JSD to witness the violations of P divisibility.
A schematic representation of such dynamics is shown in Fig.~\ref{fig:counter-representation}.

{The violation of P divisibility takes place  for $t>T_{\text{nM}}$, with $T_{\text{nM}}\approx 2.2T$ as \eqref{eq:ph-cov-P-div} is violated, and is due to positivity of the time derivative of $\eta_\perp$, in turn leading to negativity of the second condition appearing in \eqref{eq:ph-cov-P-div}. This behavior is shown in  Fig.~\ref{fig:counter-functions-rates}. Such violation corresponds to a revival in the coherences, without a corresponding revival in the population, thus building on a genuine quantum effect.}
The fact that memory effects are due to a unital feature of the map, and not to the translation $\kappa_z$, implies that $\mathscr N^D(\Phi)>0$, which, in turn, leads to $\mathscr N^\Delta(\Phi)>0$.

\begin{figure}[t]
  \includegraphics[width=\linewidth]{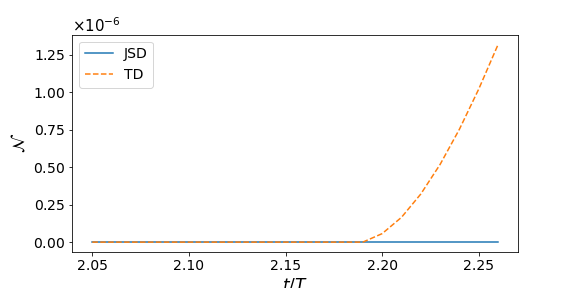}
  \caption{Non-Markovianity measure for the JSD (solid line) and for the TD (dashed line) for the considered counterexample as a function of time.
  Clearly, $\mathscr N^D(\Phi)>0$, since a revival in the TD is witnessed. The JSD, on the other hand is always a monotonic function and hence $\mathscr N^J(\Phi)=0$.
  }
  \label{fig:counter-N}
\end{figure}

On the other hand, we evaluated numerically that $\mathscr N^J(\Phi)=0$, as it can be seen in Fig.~\ref{fig:counter-N}.
The numerical analysis has been performed considering all the possible initial pairs of states on the Bloch sphere, studying their time evolution and evaluating any eventual revival of the JSD, but none has been found.
Therefore, we can conclude that the measure of \nm arising from the JSD is strictly weaker than the one arising from the Helstrom matrix.
Or, in other words, there exists non-P divisible dynamics leading to a zero measure of \nm.

Nevertheless, this fact is also true for the TD.
In order to be able to capture any violation of P divisibility as a revival of some quantifier, one has to generalise the TD to ensembles, introducing a bias parameter.
One might wonder if something similar also happens for the JSD: if we generalise it to ensembles as the Holevo skew divergence \eqref{eq:chi} or as the quantum skew divergence \eqref{eq:tre}, could we be able to witness all the violations of P divisibility?
Unfortunately, the answer to this question is no.
In fact, considering the same counterexample, one has again $\mathscr N^{K_\mu}(\Phi) = \mathscr N^{S_\mu}(\Phi)=0$.
This fact actually comes unsurprisingly: for the TD, the generalisation to ensembles breaks the translational symmetry and makes us able to detect violations of P divisibility due to the translational components of the dynamics; for the JSD, on the other hand, there is no symmetry to break and thus generalising it to ensembles does not lead to any qualitative difference.
In particular, one can consider a time evolution $\tilde \Phi$ such that $\mathscr N^{J}(\tilde\Phi)>0$, while  $\mathscr N^{D}(\tilde\Phi)=0$, as shown in \cite{Megier2021a} considering a different phase covariant model.

A natural question is what additional constraints $d$ has to obey in order to have a measure of \nm equivalent to $\mathscr N^\Delta$. 
Building on our counterexample,
a necessary condition that $d$ must obey is naturally the existence of a strictly non-contractive pair of states for any non-positive map $\Lambda$.  A second condition, crucial for entropic distinguishability quantifiers, is the relation between the non-contractivity domain $\mathcal{NCD}_{\Lambda,d}$, depending on both the map and the quantifier, and the positivity domain $\mathcal{PD}_{\Lambda}$ of the map. As we have shown, if $\mathcal{NCD}_{\Lambda,d}$ is a proper subset of $\mathcal{PD}_{\Lambda}$, detection of violation of divisibility after a time $s$ can fail for maps whose image at time $s$, by necessity within the positivity domain, is strictly outside the non-contractivity domain. This is exactly the feature we exploited to provide the counterexample.

%Thus, 

%\begin{figure}[t]
%  \centering
%  \includegraphics[width=\linewidth]{phi_dists_N}
%  \caption{Non-Markovianity for the TD and for the JSD for the counterexample.
%    On the left panel, is plotted the behaviour in time of the TD and the JSD for the optimal pairs, which are represented by antipodal vectors on the equator of the Bloch sphere.
%  Clearly, the TD has a revival, while the JSD is monotonic and thus $\mathscr N^D(\Phi)>0$, while $\mathscr N^J(\Phi)=0$, as it is evident from the right panel, showing the two measures of \nm as a function of time.}
%  \label{fig:counter-N}
%\end{figure}

\section{Conclusions and outlook}
\label{sec:conclusions-outlook}
In this work, we have compared different measures for the degree of non-Markovianity in the dynamics of open systems based on distinguishability quantifiers between quantum states. In particular, we have provided evidence that the measure based on the trace norm of the Helstrom matrix is strictly stronger than all of the measures of \nm based on entropic distinguishability quantifiers, as well as stronger than the measure based on the trace distance, which is neither stronger nor weaker than the entropic ones. This is our central result. It means that the value of the measure based on the trace norm of the Helstrom matrix associated to a dynamical map $\Phi$, namely $\mathscr N^\Delta(\Phi)$, is greater than zero whenever this happens for the measures associated to entropic distinguishability quantifiers or to the trace distance, while the reverse is not true as we have demonstrated here. Thus, we can conclude that the different distinguishability measures exhibit a quite different performance in the detection of non-positive maps, which is surprising in view of similar physical interpretations outlined in Secs.~\ref{sec:trace-dist-helstr} and \ref{sec:jensen-shannon-skew}. 

This result has been obtained considering the explicit expression of a qubit dynamics which is not P divisible, so that it is non-Markovian according to $\mathscr N^\Delta$, though strong numerical evidence shows that the associated entropic non-Markovianity measure $\mathscr N^J$, based on the Jensen-Shannon divergence as distinguishability quantifier, is equal to zero.  A purely analytic proof of this property, in the present or in a different counterexample, would provide further insights on the relationship between positivity of a map and its contractivity property with respect to entropic distinguishability quantifiers obtained from the quantum relative entropy, such as the Jensen-Shannon divergence.  

Moreover, it might be very interesting to clarify if the Helstrom based quantifier is unique, or whether other distinguishability quantifiers are equivalent to it as
measures for quantum non-Markovianity.

\acknowledgments
The authors thank Andrea Smirne for many fruitful discussions
and careful reading of the manuscript.

\bibliography{bibliographyPdiv}

\end{document}